\documentclass[prl,aps,superscriptaddress,preprint,floatfix,nofootinbib]{revtex4}

\usepackage{amsmath,amsfonts,amssymb}
\usepackage{graphicx}
\usepackage{epstopdf}
\usepackage{float}

\newcommand{\bs}[1]{\boldsymbol{#1}}

\def\3{2.8in}    
\def\2{2.5in}
\def\4{3.0in}

\def \beq {\begin{equation}}
\def \eeq {\end{equation}}

\begin{document}

\title{Magnetic and noncentrosymmetric Weyl fermion semimetals in the RAlGe family of compounds (R=rare earth)}
\author{Guoqing Chang$^*$}\affiliation{Centre for Advanced 2D Materials and Graphene Research Centre National University of Singapore, 6 Science Drive 2, Singapore 117546}\affiliation{Department of Physics, National University of Singapore, 2 Science Drive 3, Singapore 117542}
\author{Bahadur Singh$^*$}\affiliation{Centre for Advanced 2D Materials and Graphene Research Centre National University of Singapore, 6 Science Drive 2, Singapore 117546}\affiliation{Department of Physics, National University of Singapore, 2 Science Drive 3, Singapore 117542}

\author{Su-Yang Xu$^*$$^{\dag}$}\affiliation {Laboratory for Topological Quantum Matter and Spectroscopy (B7), Department of Physics, Princeton University, Princeton, New Jersey 08544, USA}

\author{Guang Bian}\affiliation {Laboratory for Topological Quantum Matter and Spectroscopy (B7), Department of Physics, Princeton University, Princeton, New Jersey 08544, USA}
\affiliation{Department  of  Physics  and  Astronomy, University  of  Missouri,  Columbia,  Missouri  65211,  USA}

\author{Shin-Ming Huang}
\affiliation{Department of Physics, National Sun Yat-sen University, Kaohsiung 804, Taiwan}
\author{Chuang-Han Hsu}\affiliation{Centre for Advanced 2D Materials and Graphene Research Centre National University of Singapore, 6 Science Drive 2, Singapore 117546}
\affiliation{Department of Physics, National University of Singapore, 2 Science Drive 3, Singapore 117542}

\author{Ilya Belopolski}\affiliation {Laboratory for Topological Quantum Matter and Spectroscopy (B7), Department of Physics, Princeton University, Princeton, New Jersey 08544, USA}

\author{Nasser Alidoust}\affiliation {Laboratory for Topological Quantum Matter and Spectroscopy (B7), Department of Physics, Princeton University, Princeton, New Jersey 08544, USA}
\author{Daniel S. Sanchez}\affiliation {Laboratory for Topological Quantum Matter and Spectroscopy (B7), Department of Physics, Princeton University, Princeton, New Jersey 08544, USA}
\author{Hao Zheng}\affiliation {Laboratory for Topological Quantum Matter and Spectroscopy (B7), Department of Physics, Princeton University, Princeton, New Jersey 08544, USA}
\affiliation {School of Physics and Astronomy, Shanghai Jiao Tong Univeristy, 200240 shanghai, China.}

\author{Hong Lu}\affiliation{International Center for Quantum Materials, School of Physics, Peking University, China}
\author{Xiao Zhang}\affiliation{International Center for Quantum Materials, School of Physics, Peking University, China}
\author{Yi Bian}\affiliation{International Center for Quantum Materials, School of Physics, Peking University, China}

\author{Tay-Rong Chang}
\affiliation{Department of Physics, National Tsing Hua University, Hsinchu 30013, Taiwan}
\affiliation{Department of Physics, National Cheng Kung University, Tainan 701, Taiwan}

\author{Horng-Tay Jeng}
\affiliation{Department of Physics, National Tsing Hua University, Hsinchu 30013, Taiwan}
\affiliation{Institute of Physics, Academia Sinica, Taipei 11529, Taiwan}

\author{Arun Bansil}
\affiliation{Department of Physics, Northeastern University, Boston, Massachusetts 02115, USA}

\author{Han Hsu}
\affiliation{Department of Physics, National Central University, Jhongli City, Taoyuan 32001, Taiwan}

\author{Shuang Jia}
\affiliation{International Center for Quantum Materials, School of Physics, Peking University, China}\affiliation{Collaborative Innovation Center of Quantum Matter, Beijing,100871, China}

\author{Titus Neupert}\affiliation {Princeton Center for Theoretical Science, Princeton University, Princeton, New Jersey 08544, USA}
\affiliation{Department of Physics, University of Zurich, Winterthurerstrasse 190, 8057 Zurich, Switzerland}

\author{Hsin Lin$^{\dag}$}
\affiliation{Centre for Advanced 2D Materials and Graphene Research Centre National University of Singapore, 6 Science Drive 2, Singapore 117546}
\affiliation{Department of Physics, National University of Singapore, 2 Science Drive 3, Singapore 117542}
\affiliation{Institute of Physics, Academia Sinica, Taipei 11529, Taiwan}

\author{M. Zahid Hasan $^{\dag}$ \footnote{Corresponding authors (emails): suyangxu@princeton.edu, nilnish@gmail.com, mzhasan@princeton.edu }}\affiliation {Laboratory for Topological Quantum Matter and Spectroscopy (B7), Department of Physics, Princeton University, Princeton, New Jersey 08544, USA}
\affiliation{Princeton Institute for Science and Technology of Materials, Princeton University, Princeton, New Jersey, 08544, USA}

\begin{abstract}
Weyl semimetals are novel topological conductors that host Weyl fermions as emergent quasiparticles. In this paper, we propose a new type of Weyl semimetal state that breaks both time-reversal symmetry and inversion-symmetry in the RAlGe (R=Rare earth) family. Compared to previous predictions of magnetic Weyl semimetal candidates, the prediction of Weyl nodes in RAlGe are more robust and less dependent on the details of the magnetism, because the Weyl nodes are already generated by the inversion breaking and the ferromagnetism acts as a simple Zeeman coupling that shifts the Weyl nodes in $k$ space. Moreover, RAlGe offers remarkable tunability, which covers all varieties of Weyl semimetals including type-I, type-II, inversion-breaking and time-reversal breaking, depending on a suitable choice of the rare earth elements. Further, the unique noncentrosymmetric and ferromagnetic Weyl semimetal state in RAlGe enables the generation of spin-currents.
\end{abstract}

\maketitle
Finding new quantum materials with useful properties is one of the frontiers of modern condensed matter physics and material science \cite{Hasan2010, Qi2011, rev1, rev2,TI_book_2015,RevBansil}. The recent realization of nonmagnetic Weyl semimetal state in the TaAs class of materials \cite{Weyl,Wan2011,Burkov2011,Huang2015, Weng2015, Hasan_TaAs,MIT_Weyl,TaAs_Ding,ARPES-NbAs, ARPES-TaP,TaAs_YL,TaAs_IOPbulk,TaAs_STM1,TaAs_STM2,nielsen1983adler,Nonlocal,Fermi arc_1,Fermi arc_2,Response}, has attracted significant attention.  Further transport measurements have revealed unconventional magnetic and optical responses of TaAs family \cite{TaAs_photocurrent,TaAs_Mag, IOP_NMR,CL_NMR}. Despite recent advances of topological semimetals in both theory\cite{WT-Weyl,Kramers,MT-Weyl, WMTe2,NewFermion,BensPaper1,trip1,trip2,Nexus,RhSi,CoSi,Nodalchain,Hopf-link,Q_photo_Cur} and experiment \cite{TaAs_photocurrent,TaAs_Mag, IOP_NMR,CL_NMR,LAG,WTe2_NM,WTe2_NP,WT-ARPES-4,WTe2_PRX,WTe2_STM,MOP,PbTaSe2}  ,  the ferromagnetic Weyl semimetal \cite{AH, Trivedi,Wan2011, Burkov2011, HgCrSe,YbMnBi2, Fisher, Ran} has not been realised in experiments. A key issue is that first-principles band structure calculations on these magnetic materials (e.g. iridates \cite{Wan2011} and HgCr$_2$Se$_4$ \cite{HgCrSe}) are quite challenging. For example, the all-in (all-out) magnetic structure in iridates appeared to be complicated to verify in experiments \cite{ Fisher} and model in first-principles calculations \cite{Wan2011, Ran}. Also, for many magnetic materials such as HgCr$_2$Se$_4$ the magnetic band structure may be very sensitive to the details of the magnetism. As a result, the first-principles prediction of Weyl nodes in magnetic compounds is not as robust as that of in nonmagnetic compounds such as TaAs \cite{Weng2015, Huang2015}. Here, we propose a new strategy to search for magnetic Weyl semimetals. Taking advantage of the Weyl nodes generated by inversion-symmetry breaking in the nonmagnetic compound LaAlGe \cite{LAG, LAG_Structure}, we present a new type of magnetic Weyl semimetal in its iso-structural sister compounds CeAlGe and PrAlGe \cite{CAG_Structure,PAG_Structure} that are ferromagnetic \cite{CeMag, RAS_Structure, Mag}. We show that the ferromagnetism in RAlGe can be more reliably modeled in first-principles calculation as it is found to not completely change the band structure. Rather, it acts as a Zeeman coupling and splits the spin-up and spin-down bands, which shifts the Weyl nodes in $k$ space to break time-reversal symmetry. For these reasons, the prediction of Weyl nodes in RAlGe are less dependent on the details of the magnetism. Moreover, we show that the RAlGe family offers remarkable tunability, where type-I, type-II \cite{WT-Weyl}, inversion breaking, and time-reversal breaking types of Weyl semimetal states are all available. Further, as recently predicted in theory \cite{Wang}, the time-reversal and inversion breaking Weyl semimetals can uniquely induce a quantum spin current without a concomitant charge current. In addition, while a noncentrosymmetric (magnetic) Weyl semimetal is an intermediate phase between a trivial insulator and a 3D topological insulator (3D stacked Chern insulator) state \cite{Murakami2007}, here, with both symmetries broken, the phase diagram may be even richer. This rich phase diagram may be potentially explored via doping or chemical substitution.

\begin{figure}[t]
\includegraphics[width=85mm]{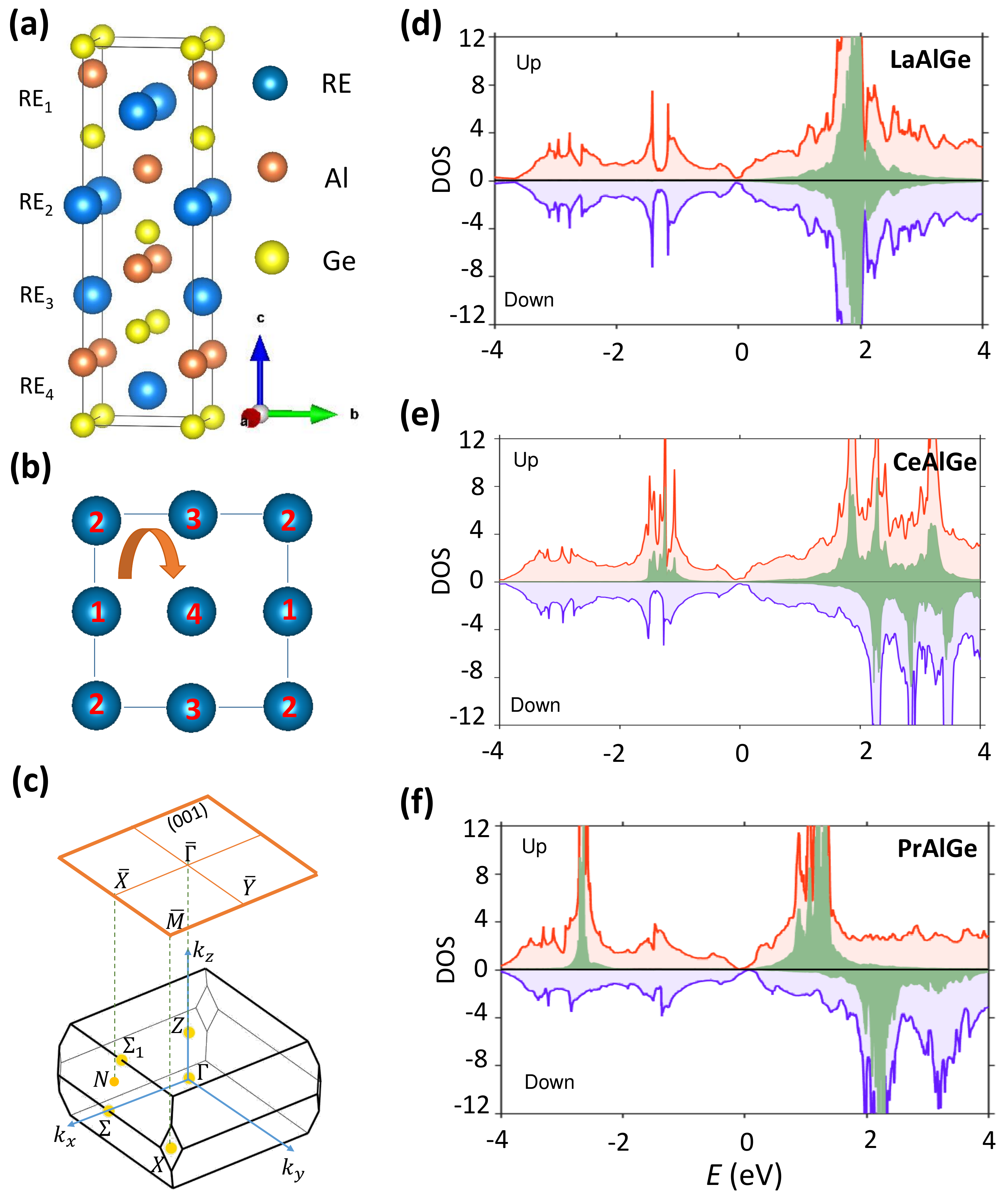}
\caption{{Lattice structure, Brillouin zone and density of state (DOS) of RAlGe (R=La, Ce, and Pr).} ({a}) Body-centered tetragonal structure of RAlGe, with space group \textit{I}$4_{1}$\textit{md} (109). The structure consists of stacks of rare earth element (RE), Al, and Ge layers and along the (001) direction each layer consists of only one type of elements. ({b}) The schematic of RE atomic layer showing the skew axis along the c axis. ({c}) The bulk and (001) surface Brillouin zone (BZ). ({d-f}) First-principles DOS of LaAlGe ({d}), CeAlGe ({e}) and PrAlGe ({f}). The partial DOS for spin-up and spin-down states are plotted in red and violet colors, respectively. The DOS from localized $f$ orbitals are drawn in green color.}
\label{Fig1}
\end{figure}



\begin{figure}[t]
\includegraphics[width=88mm]{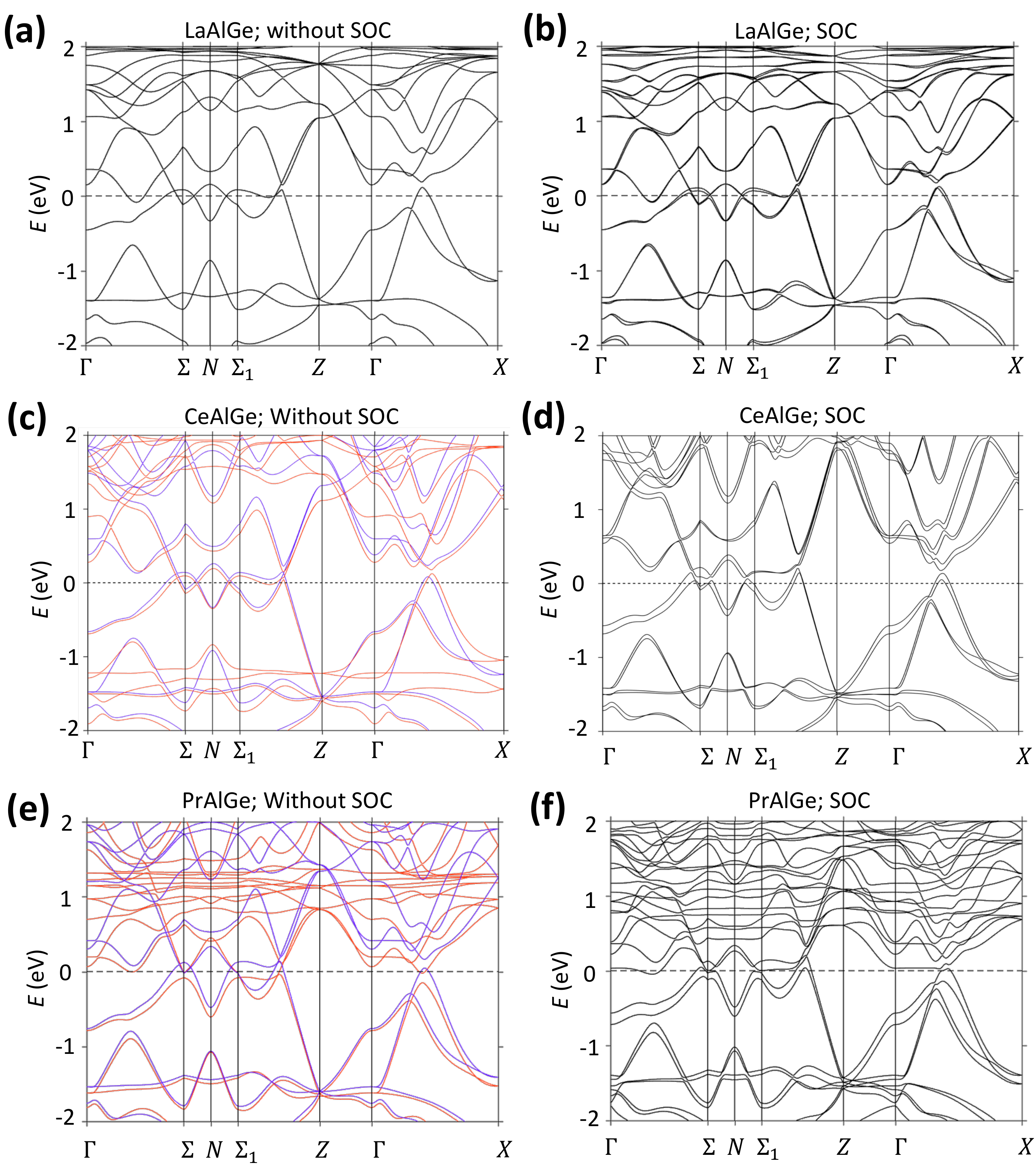}
\caption{
{ First-principles band structure of RAlGe (R=La, Ce, and Pr).} ({a,b}) Calculated bulk band structure of LaAlGe without and with the inclusion of spin-orbit coupling. ({c,d})  Bulk band structure of CeAlGe without and with the inclusion of spin-orbit coupling. In {c} the bands of spin-up and spin-down states are plotted in red and violet colors, respectively. ({e,f}) same as ({c,d}) but for PrAlGe.}
\label{Fig2}
\end{figure}

 RAlGe crystallizes in a body-centered tetragonal Bravais lattice with a $\mathcal{I}$-breaking space group $I4_1md$ (109) \cite{LAG_Structure, CAG_Structure, PAG_Structure} (Figures.~\ref{Fig1}(a,b)). Our results show that LaAlGe is nonmagnetic, whereas CeAlGe and PrAlGe are ferromagnetic with  their magnetization easy axse along the a and c directions, representively.  For CeAlGe, the calculated magnetic moment is 1 $\mu_{B}$ per Ce atom and the experimental measured value is  0.94 $\mu_{B}$ \cite{CeMag}. For PrAlGe, the calculated magnetic moment is 2 $\mu_{B}$ per Pr atom whereas the experimental value has not been reported in literature. Figs.~\ref{Fig1}(d,e,f) show the calculated density of states (DOS) without spin-orbit coupling (SOC) for RAlGe. The DOS of the majority and minority spin states are colored in red and blue. It can be seen clearly that in LaAlGe the DOS of the two spins are equal, consistent with its nonmagnetic nature. In contrast, an imbalance between the DOS of the majority and minority spin states is seen in CeAlGe and PrAlGe, suggesting a ferromagnetic ground state in agreement with the experimental finding \cite{RAS_Structure, Mag}. The green shaded areas are the DOS of $f$electrons. The electronic configuration of La atom is [Xe]$6s^25d^14f^0$, meaning that all $f$ orbitals are empty. Indeed, Fig.~\ref{Fig1}(d) shows that the all $f$ electrons are in the conduction bands. On the other hand, a Ce (or Pr) atom should have 1 (or 2) electrons occupying the $f$ orbitals. As a result, we see some $f$ bands below the Fermi level  in Figs.~\ref{Fig1}(e,f). Moreover, our calculations (Figs.~\ref{Fig1}(e,f)) show that the occupied $f$ electron states in CeAlGe and PrAlGe are clearly spin-polarized. These results suggest that the ferromagnetic coupling between the $f$ electrons' local moments lead to ferromagnetism in CeAlGe and PrAlGe, which, in turn, makes the conduction electrons ($s,p,d$ orbitals) near the Fermi level also spin-polarized. Furthermore, our band structure calculations without SOC (Figs.~\ref{Fig2}(c,e)) clearly show a spin splitting in the electronic states. These results confirm our conceptual picture: The ferromagnetism arises from the ordering of the $f$ electrons' local moments. These local moments serve as an effective Zeeman field and make the conduction ($s,p,d$ orbital) bands spin polarized. We highlight the fact that the ferromagnetism can be treated as a Zeeman coupling and does not completely change the band structure at low-energy. In the presence of SOC, the spin-up and spin-down states are further mixed by Rashba/Dresselhaus interactions due to the lack of inversion symmetry, making spin not a good quantum number. Thus, we do not color code the bands in Figs.~\ref{Fig2}(d,f).


\begin{figure}[t]
\includegraphics[width=88mm]{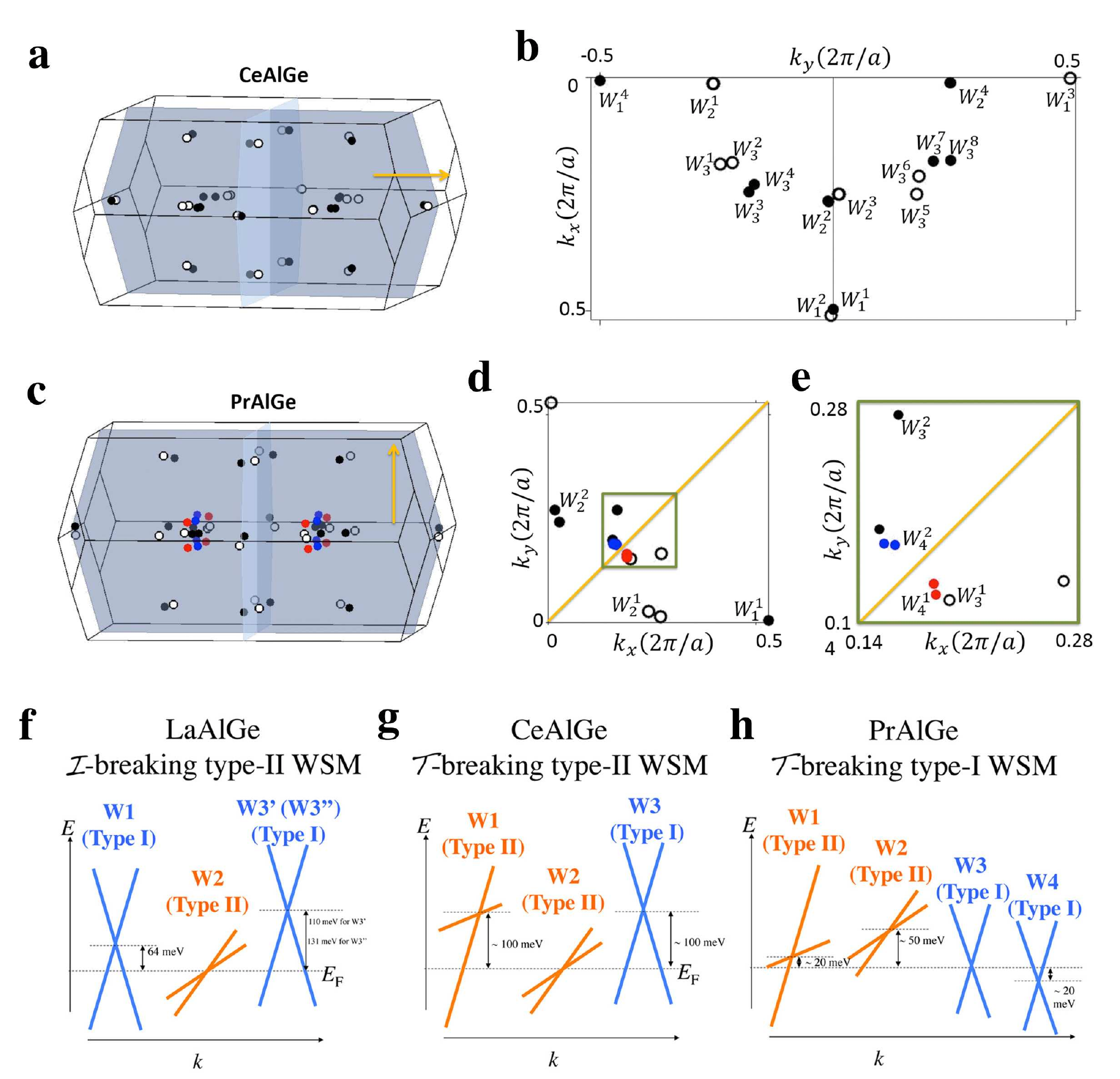}
\caption{
{Weyl fermions in LaAlGe, CeAlGe, and PrAlGe.}
 ({a}), ({c}) Weyl nodes (denoted by W3) in the first BZ of CeAlGe and PrAlGe with SOC. The arrows indicate the magnetization orientation. The red and blue dots denote the new Weyl nodes (W4) generated by the magnetization of $f$ orbitals of Pr. ({b}) Projection of the Weyl nodes on the (001) surface Brillouin zone (SBZ) of CeAlGe. The configuration of the other half of the SBZ can be obtained by considering mirror symmetry.  ({d}), ({e}) Same as {b} but for PrAlGe. ({f-h}) Schematic illustrations of the band dispersion of Weyl fermions in LaAlGe, CeAlGe, and PrAlGe, respectively.}
 \label{Fig3}
\end{figure}

In order to explain the Weyl nodes in CeAlGe and PrAlGe, we start from the nonmagnetic compound LaAlGe. In the absence of spin-orbit coupling, the crossing between conduction and valence bands yields four nodal lines, on the $k_x=0$ and $k_y=0$ mirror planes and also 4 pairs of (spinless) Weyl nodes on the $k_z=0$ plane, which we denoted as W3 \cite{LAG}. Upon the inclusion of the spin-orbit coupling, the nodal lines are gapped out and 24 Weyl nodes emerge in the vicinity. We refer to the 8 Weyl nodes located on the $k_z=0$ plane as W1 and the remaining 16 Weyl nodes away from this plane as W2 \cite{LAG}. Moreover, each W3 (spinless) Weyl node splits into two (spinful) Weyl nodes of the same chirality, which we call W3' and W3'' \cite{LAG}. Hence, in total there are 40 Weyl nodes for LaAlGe \cite{LAG}.

We now turn to the Weyl semimetal states in CeAlGe and PrAlGe. We conceptually consider a temperature dependent evolution. Starting at a higher temperature above the Curie transition, we expect the Ce(Pr)AlGe sample to already become a Weyl semimetal because of the broken space-inversion symmetry with 40 Weyl nodes as in LaAlGe. Now we lower the temperature below the Curie temperature, the effect of the ferromagnetism in CeAlGe and PrAlGe can be understood qualitatively as a Zeeman coupling to the conduction electron states. To the lowest order, we expect that this will shift the Weyl nodes in a way that their momentum space configuration reflects the time-reversal symmetry breaking. We use this picture to understand the calculated results of the Weyl nodes configuration of these two compounds. In CeAlGe, indeed, we found that the Weyl nodes are still the W1, W2, and W3 as in LaAlGe (Fig.~\ref{Fig3}(a, b)). The difference is that they are shifted away from the original location due to magnetism. In LaAlGe all W1 nodes can be related by symmetry operations. However, in CeAlGe, the inclusion of a magnetization along $a$ direction gives rise to 4 inequivalent W1 Weyl nodes. They have different momentum space locations and energies. Similarly, there are now 4 inequivalent W2 and 8 inequivalent W3 Weyl nodes in CeAlGe because of the reduction of symmetries by the inclusion of the magnetization. In PrAlGe, the magnetization along the $c$ axis leads to 1 inequivalent W1, 2 inequivalent W2 and 2 inequivalent W3 Weyl nodes. In addition, we find that the inclusion of ferromagnetization in PrAlGe may introduce new Weyl nodes, which we denote as the W4 nodes (Figs.~\ref{Fig3}(c-e)). The chiral charges of Weyl points in RAlGe are determined by the net Berry flux passing through the 2D manifold that enclosing Weyl fermions\cite{Wan2011}.

In order to understand how the Weyl nodes are shifted by the magnetization, we discuss the symmetry constraints in the presence of the ferromagnetic order in CeAlGe and PrAlGe.

In CeAlGe, the magnetization is oriented along $a$ axis. Both $\mathcal{T}$ and $C_2$ reverse this in-plane magnetization. However, their product $C_2\mathcal{T}$ is still a symmetry of the magnetic system and the same is true for $M_x$. Thus, all symmetry-nonequivalent Weyl nodes are found in the $k_x>0$ part of the BZ depicted in Fig.~\ref{Fig3}(b). Due to $C_2\mathcal{T}$, all W1- and W3-derived Weyl nodes are still pinned to $k_z=0$ and the W2-derived nodes are found in $\pm k_z$ pairs. It is also interesting to notice that the movement of the all Weyl nodes in the vicinity of the $M_y$ ($//k_x$) mirror plane (W$_1^1$, W$_1^2$, W$_2^2$, and W$_2^3$), is much more significant than those of the $M_x$ ($//k_y$) mirror plane (W$_1^3$, W$_1^4$, W$_2^1$, and W$_2^4$). This phenomenon is also symmetry related. Specifically, the Weyl nodes near the  $M_x$ ($//k_y$) mirror plane are roughly stationary upon magnetization because the symmetries $C_2\mathcal{T}$ and $M_x$ are the only constraints to the effective Hamiltonian near the $M_x$ mirror plane \cite{Weng2015, Huang2015}. However, this term turns out to be only relevant for the energy but not for the position of the Weyl nodes. The detailed information of the Weyl nodes including the momentum space locations, the energies and the type, is shown in the Supplemental Material, Sec. C \cite{SM} .

In PrAlGe, the magnetization is oriented along $z$. Both $\mathcal{T}$ and any of the mirror and glide mirror symmetries reverse this magnetization. However, their products, e.g., $\mathcal{T}M_x$, are still a symmetry of the magnetic system. Furthermore, $C_{2z}$ and $\bar{C}_{4z}$ are preserved by the magnetization. Thus, all symmetry-nonequivalent Weyl nodes are found in a quadrant of the BZ depicted in Fig.~\ref{Fig3}(d). The main difference to LaAlGe is that $C_{2z}\mathcal{T}$ symmetry is broken. As a result, we expect the W1 and W3 Weyl nodes to move along the $k_z$ direction and become no longer pinned to $k_z=0$. On the other hand, the W2  Weyl nodes are expected to stop appearing in $\pm k_z$ pairs.

\begin{figure}[t]
\includegraphics[width=88mm]{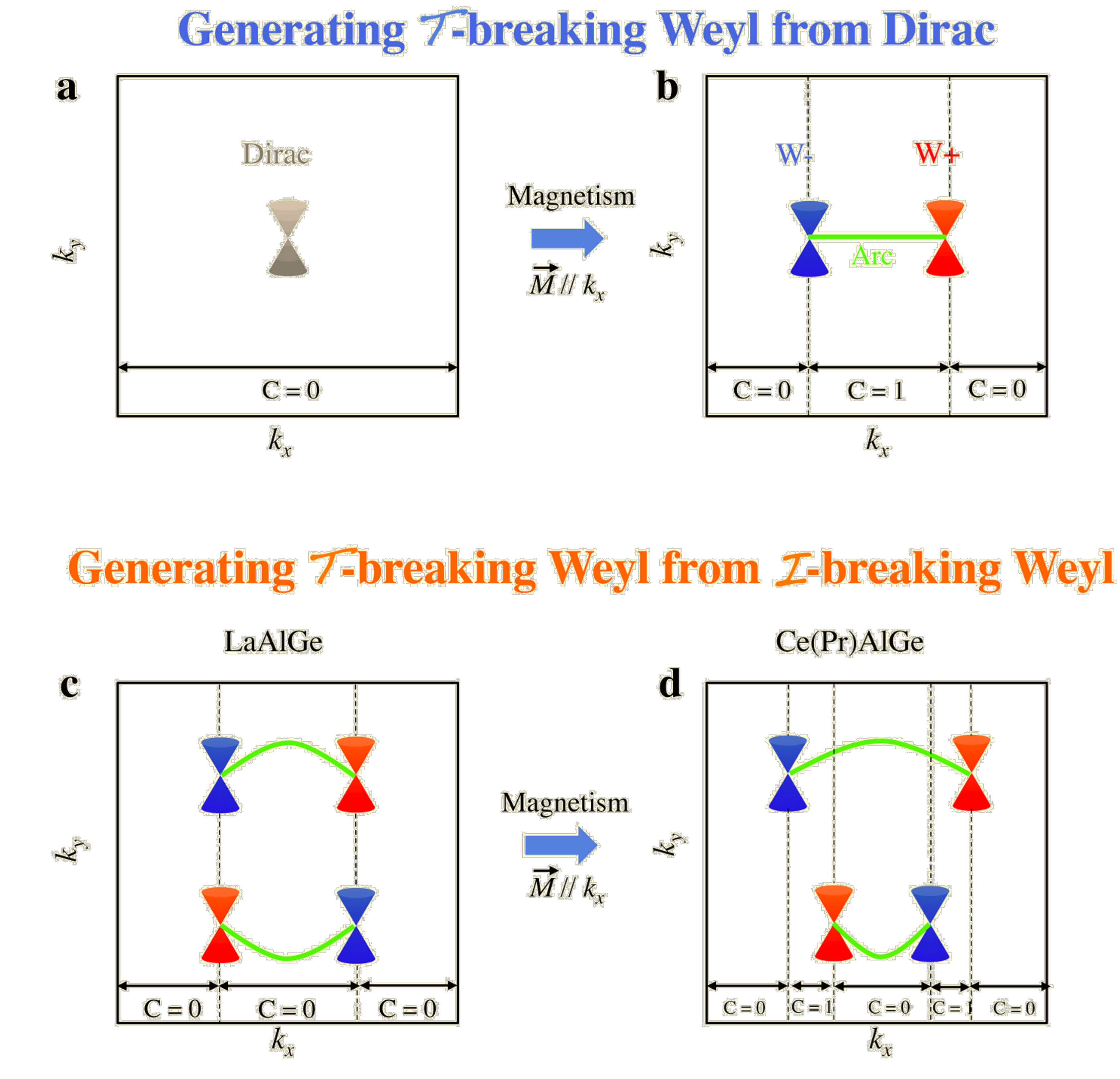}
\caption{ { A new route to generating magnetic Weyl fermions.} ({a,b,}) The Dirac node in a Dirac semimetal splits into a pair of Weyl nodes upon the inclusion of a magnetic field. The Weyl nodes are connected by a Fermi arc surface state. ({c,d,}) Two pairs of Weyl nodes are present in a inversion-breaking Weyl semimetal. The Weyl nodes are shifted in momentum space upon the inclusion of a magnetic field, generating regions in $ \textit{\textbf{k}}$ space with a non-zero Chern number.}
\label{Fig4}
\end{figure}


Our theoretical discovery of CeAlGe and PrAlGe reveals a new route to realizing $\mathcal{T}$-breaking Weyl fermions. The traditional and commonly accepted proposal for realizing $\mathcal{T}$-breaking Weyl fermions is to break time-reversal symmetry of a 3D Dirac fermion system, as shown in Figs.~\ref{Fig4}(a,b). In this way, the Weyl fermions actually arise from the breaking of time-reversal symmetry. We can qualitatively understand the anomalous Hall effect. As shown in Fig.~\ref{Fig4}(b), we consider the Chern number of a series of ($k_y, k_z$) 2D slices at different $k_x$-intercepts. Any slice between the left boundary of the BZ and the first dotted line has a Chern number of 0. As we continue sweeping the ($k_y, k_z$) slice to the right, we pass through the blue Weyl node and therefore the Chern number changes by 1. Consequently, a slice between the first dotted line and the second dotted line has a Chern number of 1. Then we pass the red Weyl node and the Chern number of a slice between the second dotted line and the right boundary of the BZ is 0. As a result, the Chern number averaged over all $k_z$ in the BZ is nonzero, which demonstrates the existence of an anomalous Hall effect. It can be checked that this simple consideration carries over to CeAlGe and PrAlGe, despite the presence of additional symmetries, and an anomalous Hall effect is expected in the plane perpendicular to the respective magnetization. By contrast, in the Dirac semimetal case in Fig.~\ref{Fig4}(a), the Chern number of any slice is zero, consistent with the fact that Dirac semimetals do not show anomalous Hall conductance. In terms of experimental realization, this proposal means that one needs to introduce magnetism to a Dirac semimetal system such as TlBi(S$_{1-x}$Se$_x$)$_2$, Cd$_3$As$_2$ and Na$_3$Bi. Since these materials are nonmagnetic, one will need to dope the system with magnetic dopants, which has been proven to be quite difficult. It is also challenging to systematically study the band structure of the magnetically doped system through first-principles calculations. 

We now elaborate on our new route to realizing $\mathcal{T}$-breaking Weyl fermions as demonstrated in CeAlGe and PrAlGe. Rather than starting from a Dirac semimetal, we start from a space-inversion ($\mathcal{I}$) breaking Weyl semimetal. As schematically shown in Fig.~\ref{Fig4}(c), two pairs of Weyl nodes are generated by the breaking of space inversion symmetry. In this case, magnetization is only responsible for shifting the momentum space location of the Weyl nodes. We note that a $\mathcal{T}$-breaking Weyl semimetal is defined as the breaking of time-reversal symmetry in terms of the Weyl node configuration. Specifically, Weyl nodes of same chirality cannot appear at opposite momenta ($\pm{\vec{k}}$). Therefore, although in this case the Weyl fermions do not arise from ferromagnetism, the system in Fig.~\ref{Fig4}(d) still counts as a $\mathcal{T}$-breaking Weyl semimetal. This can also be seen by studying the Chern number of the ($k_y, k_z$) 2D slices. As shown in Figs.~\ref{Fig4}(c,d), introducing magnetism leads to a finite $k_x$ range at which the Chern number of the ($k_y, k_z$) 2D slice is nonzero. This also suggests the existence of anomalous Hall conductance in the system shown in Fig.~\ref{Fig4}(d). We emphasize a number of advantages of this new route. Introducing magnetism is done by going from LaAlGe to CeAlGe or PrAlGe rather than doping. This not only avoids the complicated doping processes, but also enables to systematically understand the band structure in calculations as we have achieved here. Furthermore, our results demonstrate an entirely new way to search for $\mathcal{T}$-breaking Weyl semimetals in future, i.e., to look for the iso-chemical ferromagnetic cousin compounds of an $\mathcal{I}$-breaking Weyl semimetal.

Finally, we highlight the tunability of the RAlGe family. As we have shown here, the low-energy band structures of LaAlGe, CeAlGe and PrAlGe realize the $\mathcal{I}$-breaking type-II, the $\mathcal{T}$-breaking type-II, and the $\mathcal{T}$-breaking type-I Weyl fermions. Moreover, $n$ (electron) doping can be achieved by changing the ratio between Al and Ge, i.e., RAl$_{1-x}$Ge$_{1+x}$ \cite{PAG_Structure}. In the weak disorder limit ($x\ll1$), which cannot localize the conduction electorns from the Weyl fermions \cite{Xie, Hughes}, the doping will enable one to access other Weyl nodes that are above the Fermi level (Figs.~\ref{Fig3}(f-h)). In general, RAlGe is an extremely rich system that enables one to systematically study all types of Weyl fermions in a single family.

Work at Princeton was supported by the US Department of Energy under Basic Energy Sciences (Grant No. DOE/BES DE-FG-02-05ER46200). M.Z.H. acknowledges Visiting Scientist support from Lawrence Berkeley National Laboratory, and partial support from the Gordon and Betty Moore Foundation for theoretical work. The work at the National University of Singapore was supported by the National Research Foundation, Prime Minister's Office, Singapore under its NRF fellowship (NRF Award No. NRF-NRFF2013-03). S. J.acknowledge support by National science foundation of china No.11774007 and National Basic Research Program of China Grant No. 2014CB239302. The work at Northeastern University was supported by the US Department of Energy (DOE), Office of Science, Basic Energy Sciences grant number DE-FG02-07ER46352, and benefited from Northeastern University's Advanced Scientific Computation Center (ASCC) and the NERSC supercomputing center through DOE grant number DE-AC02- 05CH11231.  T.N. acknowledge support by the Swiss National Science Foundation (grant number 200021-169061) and the ERC-StG-Neupert-757867-PARATOP, respectively. The work at the National Sun Yat-sen University was supported by the Ministry of Science and Technology in Taiwan under Grant No. MOST105-2112-M110-014-MY3. T.-R.C. and H.-T. J are supported by the Ministry of Science and Technology.  T.-R. C is  supported by National Cheng Kung University. T.-R.C. and H.-T.J. also thank the National Center for Theoretical Sciences (NCTS) for technical support. 

G.C., B.S., and S.-Y.X. contribute equally to this work.

\newpage

\clearpage

\textbf{
\begin{center}
{\large \underline{Supplementary Material}: \\Magnetic and noncentrosymmetric Weyl fermion semimetals in the RAlGe family of compounds (R=rare earth)}
\end{center}
}

\vspace{0.2cm}

\begin{center}

\end{center}

\vspace{0.25cm}

\subsection{\large SM A. Methods}

The first-principles calculations were performed within the density functional theory (DFT) framework using the projector augmented wave method \cite{dft2} as implemented in the VASP \cite{dft3} package and full-potential augmented plane-wave method as implemented in the package Wiek2k. The generalized gradient approximation (GGA) \cite{dft4} was used for the exchange-correlation effect and the Hubbard energy $U$ used in the calculation is $4$ eV. The lattice constants of RAlX were acquired from previous experimental measurements \cite{LAG_Structure, CAG_Structure, PAG_Structure}. A $\Gamma$-centered $k$-point $14 \times 14 \times 14$ mesh was used and spin-orbit coupling was included in the self-consistent cycles.

\clearpage

\clearpage

\subsection{{\large SM B. Band dispersion of Weyl quasi-particles in LaAlGe, CeAlGe, and PrAlGe. }}

\begin{figure}[t]
\includegraphics[width=17cm]{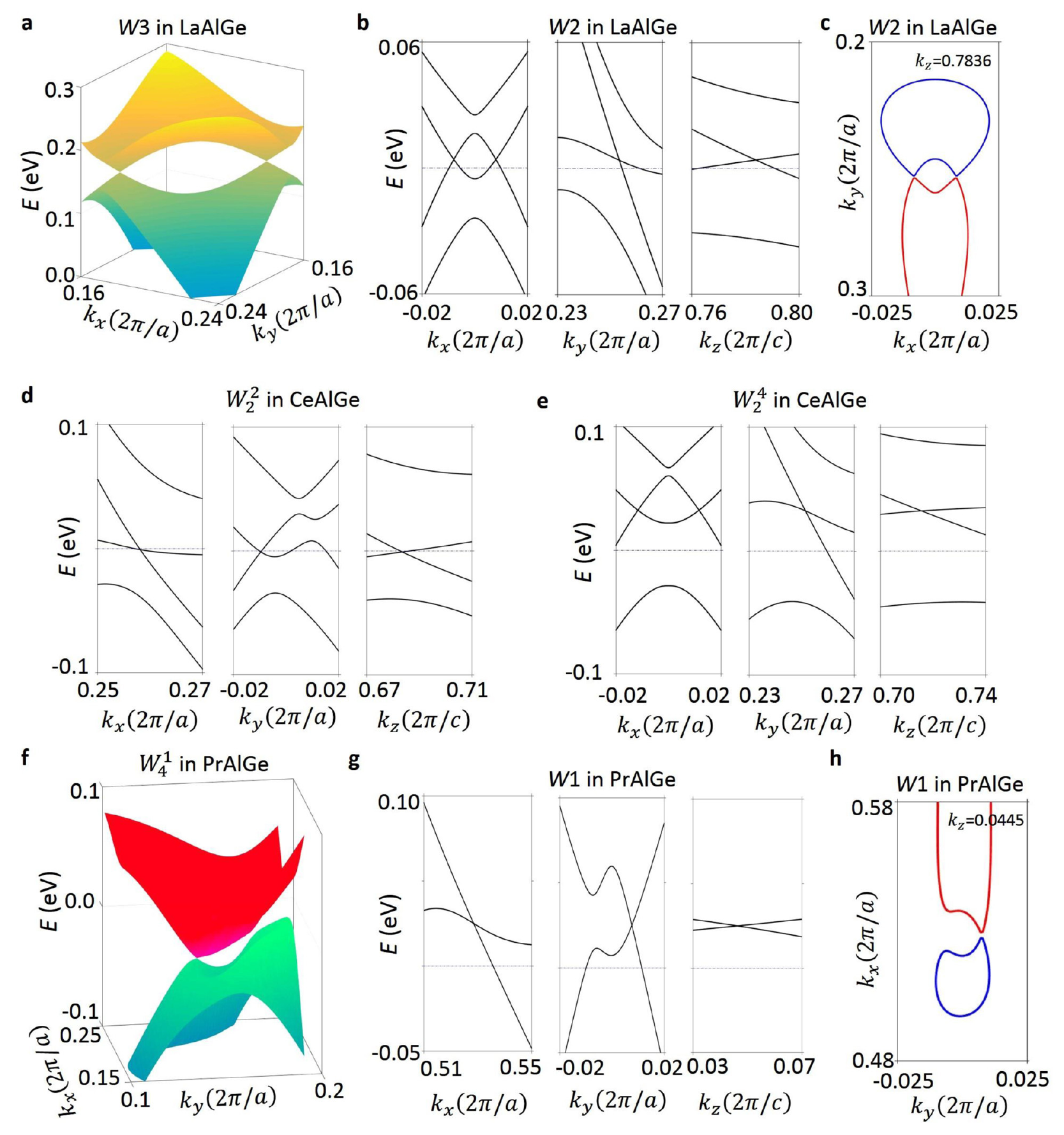}
\caption{ {Band dispersion of Weyl quasi-particles in LaAlGe, CeAlGe, and PrAlGe.} ({a}) The bulk band structure close to $W3$ Weyl nodes in LaAlGe. ({b}) The linear band dispersion along three momentum directions showing the type-II Weyl fermion nature of the $W2$ node in LaAlGe. ({c}) calculated constant energy band contour in the $k_z$=0 plane at the energy of the W2 nodes of LaAlGe. Blue lines correspond to hole-like pockets, whereas red lines indicate electron-like pockets. ({d}) Same as {b}, but for $W_{2}^{2}$ Weyl nodes in CeAlGe. ({e}) Same as {b}, but for $W_{2}^{4}$ Weyl nodes in CeAlGe. ({f}) Same as {a}, but for $W_{4}^{1}$ Weyl nodes in PrAlGe.  ({g}) Same as {b}, but for $W1$ Weyl nodes in PrAlGe.  ({h}) Same as {c}, but for $W1$ Weyl nodes in PrAlGe in the $k_z$=0.0445 plane.}

\label{Fig5}
\end{figure}

In this section, we describe the dispersion of the Weyl cones. The energy dispersions of Weyl nodes W3 and W2 in LaAlGe are plotted in Fig 5 (a) and (b) representatively. Experimental measurements of Tyep II Weyl nodes W2 in LaAlGe are consistent with our calculations in Fig S1 (b) and (c)\cite{LAG}. In CeAlGe, we show the energy dispersions of the W$_2^2$ and W$_2^4$ Weyl cones in Figs.~\ref{Fig5}(d,e). We see that the W2 are still type-II. Because the magnetization along the $a$ direction breaks the mirror symmetry $M_x$, we see in Fig 3 (b) that the two W2 Weyl nodes in the vicinity of $M_x$, i.e., W$_2^2$ and W$_2^3$, are not symmetric with respect to this mirror any more. This can also be seen in the energy dispersions near W$_2^2$ in Fig.~\ref{Fig5}(d). The W1 and W3 Weyl nodes are far away from the Fermi level and therefore not relevant to low-energy physics. W2, which are type-II, are very close to the Fermi level. Therefore, CeAlGe is a $\mathcal{T}$-breaking type-II Weyl semimetal. The Weyl fermion dispersions of PrAlGe are shown in Figs.~\ref{Fig5}(f-h). The dispersion in $(E,k_x,k_y)$ space of the newly generated W4 Weyl cone is shown in Fig.~\ref{Fig5}(f), from which we see that W4 are type-I. Figures~\ref{Fig5}(g,h) show the dispersion and Fermi surface of a W1 Weyl cone. We see that the W1 Weyl cones in PrAlGe also become type-II. We further observe that the W1 Weyl nodes are shifted away from the $k_z=0$ plane (Fig.~\ref{Fig5}(h)). In fact, the two adjacent W1 nodes move in opposite $k_z$ directions. As a result, the Fermi surface at a constant $k_z$ (Fig.~\ref{Fig5}(h)) only reveals one Weyl node. A schematic summary is shown in Fig. 3 (g). Interestingly, W2 is pushed to be well above the Fermi level. On the other hand, W3 and W4, which are of type-I, are now close to the Fermi level. Therefore, PrAlGe is a $\mathcal{T}$-breaking type-I Weyl semimetal.

\clearpage
\subsection{{\large SM C. Energy, momentum space locations, and type of Weyl nodes in LaAlGe, CeAlGe, and PrAlGe}}
\begin{center}
\begin{table}[H]
\begin{tabular}{p{3cm}p{2.5cm}p{2.5cm}p{2.5cm}p{2cm}p{2cm}}
\hline
Weyl nodes & $k_x$ ($\frac{\pi}{a}$) & $k_y$ ($\frac{\pi}{a}$) & $k_z$ ($\frac{\pi}{a}$) &  $E$ (meV) & Type\\
\hline
 $W1$ & 0.503 & 0.002 & 0.000 &  64 & I \\
$W2$ & 0.254 & 0.008 & 0.784 & 3 &  II\\
 $W3'$ & 0.216 & 0.173 & 0.000 &   110 &  I\\
 $W3''$ & 0.231 & 0.183 & 0.000 &  131 &  I\\
\hline
\end{tabular}
\caption{\label{Weyl} \textbf{Energy and momentum space locations of the Weyl nodes in LaAlGe}}


\end{table}
\end{center}

\begin{center}
\begin{table}
\begin{tabular}{p{3cm}p{2.5cm}p{2.5cm}p{2.5cm}p{2cm}p{2cm}}
\hline
Weyl nodes & $k_x$ ($\frac{\pi}{a}$) & $k_y$ ($\frac{\pi}{a}$) & $k_z$ ($\frac{\pi}{a}$) &  $E$ (meV) & Type\\
\hline
$W_{1}^1$ & 0.499 & -0.002 & 0.000 &  107 &  I \\
$W_{1}^2$ & 0.511 & -0.005 & 0.000 &  91 &  I\\
$W_{1}^3$ & 0.005 & 0.512 & 0.000 &  116 &  I\\
$W_{1}^4$ & 0.007 & -0.499 & 0.000 &  116 &  I\\
$W_{2}^1$ & 0.252 & 0.012 & 0.722 &  37 &  II \\
$W_{2}^2$ & 0.266 & -0.010 & 0.687 &  -1 &  II \\
$W_{2}^3$ & 0.011 & 0.251 & 0.716 &  32 &  II \\
$W_{2}^4$ & 0.014 & -0.258 & 0.705 &  15 &  II \\
$W_{3}^1$ & 0.185 & -0.243 & 0.000 &  112 &  I \\
$W_{3}^2$ & 0.183 & -0.217 & 0.000 &  90 &  I \\
$W_{3}^3$ & 0.246 & -0.180 & 0.000 &  107 &  I \\
$W_{3}^4$ & 0.229 & -0.169 & 0.000 &  66 &  I \\
$W_{3}^5$ & 0.250 & 0.179 & 0.000 &  76 &  I \\
$W_{3}^6$ & 0.212 & 0.184 & 0.000 &  106 &  I \\
$W_{3}^7$ & 0.180 & 0.214 & 0.000 &  96 &  I \\
$W_{3}^8$ & 0.178 & 0.252 & 0.000 &  75 &  I \\
\hline
\end{tabular}
\caption{\label{Weyl} \textbf{Energy and momentum space locations of the Weyl nodes in CeAlGe with a (100) magnetization.}}

\end{table}
\end{center}

\clearpage
\begin{center}
\begin{table}
\centering
\begin{tabular}{p{3cm}p{2.5cm}p{2.5cm}p{2.5cm}p{2cm}p{2cm}}
\hline
Weyl nodes & $k_x$ ($\frac{\pi}{a}$) & $k_y$ ($\frac{\pi}{a}$) & $k_z$ ($\frac{\pi}{a}$) &  $E$ (meV) & Type\\
\hline
$W_{1}$ & 0.530 & 0.007 & 0.045 &  25 & II \\
$W_{2}^1$ & 0.027 & 0.242 & 0.718 &  54 & II\\
$W_{2}^2$ & 0.271 & 0.016 & 0.747 &  51 & II\\
$W_{3}^1$ & 0.230 & 0.194 & 0.038 &  14 & I\\
$W_{3}^2$ & 0.161 & 0.268 & 0.037 &  -5 & I\\
$W_{4}^1$ & 0.190 & 0.157 & 0.115 &  -23 & I\\
$W_{4}^2$ & 0.164 & 0.189 & 0.136 &  -26 & I\\
\hline
\end{tabular}
\caption{\label{Weyl} \textbf{Energy and momentum space locations of the Weyl nodes in PrAlGe with a (001) magnetization.}}

\end{table}
\end{center}

\clearpage
\subsection{{\large SM D. Hamiltonian for glide mirror planes (W3 nodes)}}

In RAlGe, the W1 and W2 Weyl nodes arise from the spinless nodal lines, which is the same as TaAs. However, the W3' and W3'' Weyl nodes are due to the spinless (W3) Weyl nodes. These nodes are located near the $k_x=\pm k_y$ ($45^{\circ}$) planes. Here we provide a Hamiltonian for the glide mirror planes, which can be used to fit the band structure near these glide mirrors. On the glide mirror planes $k_x=\pm k_y$, there are two relevant symmetries: $C_2$ times time-reversal and the glide-$C_4$ times mirror. Both restrict the form of the $\bs{k}\cdot\bs{p}$ Hamiltonian on this plane.
Consider a general spinless two-band Hamiltonian
\begin{equation}
H(\bs{k})=d_1(\bs{k})\sigma_1+d_2(\bs{k})\sigma_2+d_3(\bs{k})\sigma_3,
\end{equation}
where $\bs{k}=(k_1,k_2,k_z)^{\mathsf{T}}$ with $k_1:=k_x-k_y$ and $k_2=k_x+k_y$. Let us without loss of generality consider the $k_1=0$ glide mirror plane.

The symmetries act as
\begin{equation}
C_{2}\mathcal{T}H(k_1,k_2,k_z)C_{2}\mathcal{T}^{-1}
=H(k_1,k_2,-k_z)^*
\label{eq: C2T spinless}
\end{equation}
and
\begin{equation}
C_{4M}H(k_1,k_2,k_z)C_{4M}^{-1}
=H(-k_1,k_2,k_z)
\end{equation}
where $C_{2}\mathcal{T}$ and $C_{4M}$ are unitary operators that equal the identity when raised to the second power. Since we have spinless electrons, $C_{2}\mathcal{T}=C_2$. Furthermore, the Hamiltonian should be invariant under TRS
\begin{equation}
H(k_1,k_2,k_z)
=H(-k_1,-k_2,-k_z)^*,
\label{eq: TRS spinless}
\end{equation}
which tells us that $d_2(\bs{k})$ (and only this one) must be odd in $\bs{k}$. The $C_2$ symmetry implies that
\begin{equation}
C_{2}H(k_1,k_2,k_z)C_{2}^{-1}
=H(-k_1,-k_2,k_z).
\end{equation}
From the first-principle calculation, we know that both bands of interest are in the same irreducible representation, i.e., $C_{2T}=C_{4M}=\sigma_0$.
Then Eq.~\eqref{eq: C2T spinless} implies together with Eq.~\eqref{eq: TRS spinless} that
\begin{equation}
d_2(\bs{k})=vk_z+\mathcal{O}(\bs{k}^3).
\end{equation}
The other components of the Hamiltonian are even functions and otherwise unrestricted
\begin{equation}
\begin{split}
d_1(\bs{k})=&m-u_1k_1^2-u_2k_2^2-u_3k_z^2+\mathcal{O}(\bs{k}^4),\\
d_3(\bs{k})=&m'-w_1k_1^2-w_2k_2^2-w_3k_z^2+\mathcal{O}(\bs{k}^4).
\end{split}
\end{equation}
Notice that cross-terms $k_1 k_2$, $k_1 k_z$, and $k_2 k_z$ are not symmetry-allowed, due to $C_{4M}$ and $C_{2T}$, respectively..
This Hamiltonian gives rise to two pairs of Weyl nodes on the glide mirror plane at $k_z=0$, and
\begin{equation}
k_1^2=\frac{mw_2-m'u_2}{u_1w_2-u_2w_1},
\qquad
k_2^2=\frac{mw_1-m'u_1}{u_1w_2-u_2w_1},
\end{equation}
provided the coefficients are such that the right hand sides are positive.

We now want to study the effect of spin-orbit coupling. To split the Weyl nodes, we need terms that are nonzero on the $k_z$ plane. Since under $C_2\mathcal{T}$ we have $k_z\to -k_z$ and $s_1\to -s_1$, $s_2\to -s_2$, $s_3\to +s_3$, we can add the following five terms to the Hamiltonian that are nonzero on the $k_z=0$ plane:
\begin{equation}
\sigma_0 s_3,
\quad
\sigma_1 s_3,
\quad
\sigma_2 s_1,
\quad
\sigma_2 s_2,
\quad
\sigma_3 s_3.
\end{equation}
The first term commutes with the rest of the Hamiltonian and will therefore not in itself be important for splitting the Weyl nodes. The other important symmetry is $C_{4M}=\mathrm{i}(s_2-s_1)/\sqrt{2}$ that sends $k_1\to -k_1$ and $s_1\to -s_2$, $s_2\to -s_1$, $s_3\to -s_3$. We thus add to the Hamiltonian
\begin{equation}
H_{\mathrm{SOC}}
=
m_1(\bs{k})
\sigma_0 s_3
+
m_2(\bs{k})
\sigma_1 s_3
+
m_3(k_1,k_2,k_z)
\sigma_2 s_1
-
m_3(-k_1,k_2,k_z)
\sigma_2 s_2
+
m_5(\bs{k})
\sigma_3 s_3,
\end{equation}
where $m_{1,2,5}(k_1,k_2,k_z)=-m_{1,2,5}(-k_1,k_2,k_z)$.
Interestingly, there are now two qualitatively distinct ways to split the Weyl nodes: First, if we only consider $m_3(\bs{k})=\bar{m}_3+a k_1$ and set all other $m_{1,2,5}(\bs{k})=0$, then the spectrum is
\begin{equation}
E=\pm \sqrt{d_1^2+\left(d_2\pm \sqrt{2}\sqrt{\bar{m}_3^2+a^2k_1^2}\right)^2+d_3^2}
\end{equation}
which shifts the Weyl nodes away from the $k_z=0$ plane to $k_z=\pm\sqrt2\bar{m}_3/v$ (sending $a\to0$ for simplicity). However, they have pairwise the same $k_1$ and $k_2$ values.

When $k_1$-odd terms are considered, we observe that these are competing with the $\bar{m}_3$ term and can lead to a splitting of the Weyl nodes within the $k_z=0$ plane. Obtaining a closed form for the energies when all terms are added is hard, hence we only illustrate the case by considering $m_{2,5}(\bs{k})\neq0$,$m_{1,3}=0$ . Then
\begin{equation}
E=\pm \sqrt{(d_1\pm m_2)^2+d_2^2+(d_3\pm m_5)^2},
\end{equation}
which in view of the fact that $d_2=0$ is satisfied on the $k_z=0$ plane leaves the Weyl nodes in that plane, but will move their $k_1,k_2$ position.

\clearpage
\subsection{{\large SM E. HgCr2Se4 }}
\begin{figure*}[t]
\includegraphics[width=17cm]{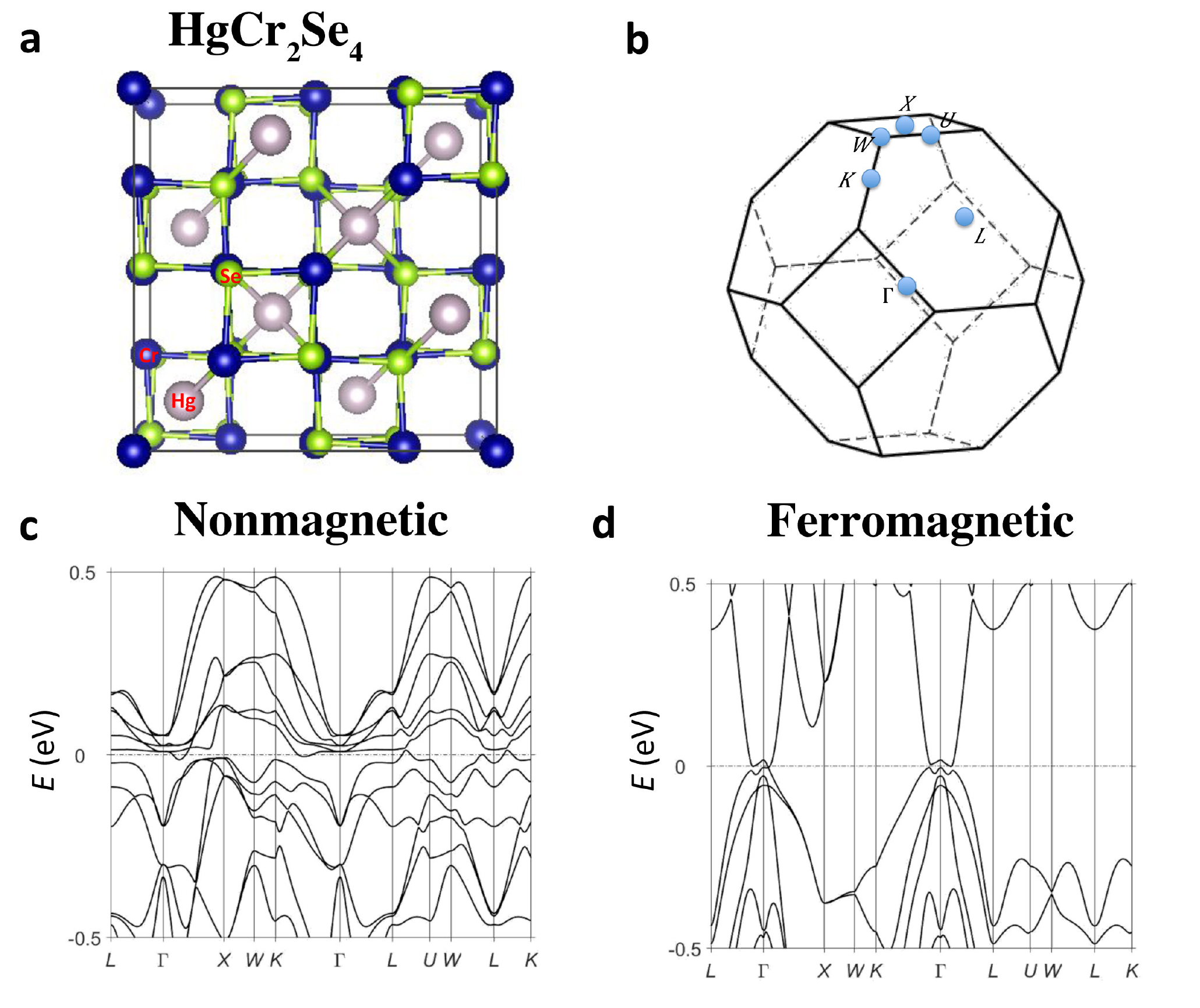}
\caption{ {HgCr2Se4.} ({a}) The crystal structure of HgCr2Se4. ({b}) The Brillouin zone of HgCr2Se4 ({c}) The bands of HgCr2Se4 without ferromagnetism. ({d}) The bands of HgCr2Se4 with ferromagnetism.}

\label{Fig6}
\end{figure*}

In this section, we discuss about the previous prediction of ferromagnetic Weyl semimetal state in HgCr2Se4 \cite{HgCrSe}.  The crystal structure and the Brillouin zone of HgCr2Se4 are plotted in Figs.~\ref{Fig6}(a,b). A key issue is that DFT band structure calculations on these magnetic materials  HgCr2Se4 are quite challenging. As shown in Figs.~\ref{Fig6}(c,d), the DFT band structures of HgCr2Se4 without (Fig.~\ref{Fig6}(c)) or with (Fig.~\ref{Fig6}(d)) ferromagnetism appears to be vastly different. Without magnetization, this system has continues gap between the conduction and valance band. After turning on ferromagnetic effect, the up and down spins split dramatically. As a result, the DFT prediction of Weyl nodes in these compounds is not as robust as that of in TaAs.

\end{document}